\documentstyle[preprint,aps,epsf]{revtex}
\tightenlines

\newcommand{\nn}{\nonumber}
\def\dfrac#1#2{\displaystyle\frac{#1}{#2}}
\newcommand{\ovl}[1]{\overline{#1}}
\newcommand{\wt}[1]{\widetilde{#1}}
\newcommand{\eq}[1]{Eq.~(\ref{#1})}
\newcommand{\eqn}[1]{(\ref{#1})}

\newcommand{\pslash}{p\kern-1ex /}
\newcommand{\lslash}{l\kern-1ex /}
\newcommand{\Dslash}{{\cal D}\kern-1.5ex /}
\newcommand{\bpsi}{\overline{\psi}}
\newcommand{\bc}{\overline{c}}
\newcommand{\tr}{{\rm tr}}

\begin{document}
\draft

\title{
{\normalsize
\begin{flushright}
UTHEP-373\\
MPI/PhT/97-70\\
November 1997\\
\end{flushright}}
One loop calculation in lattice  QCD with domain-wall quarks}
\author{$^{1,2}$Sinya Aoki 
and 
$^1$Yusuke Taniguchi
}
\address{
$^1$Institute of Physics, University of Tsukuba, Tsukuba 305, Japan \\
$^2$Max-Planck-Institut f\"ur Physik, F\"oringer Ring 6, D-80805
 M\"unchen, Germany }

\date{\today}

\maketitle

\begin{abstract}
One loop corrections to the
domain-wall quark propagator are calculated in massless QCD.
It is shown that
no additative counter term to the current quark mass
is generated in this theory,
and the wave function renormalization factor of the massless quark
is explicitly evaluated.
We also show that an analysis with a simple mean-field
approximation can explain properties of 
the massless quark in numerical simulations of QCD 
with domain-wall quarks.
\end{abstract}

\pacs{11.15Ha, 11.30Rd, 12.38Bx, 12.38.Gc}

\narrowtext

%
\section{Introduction}

The formulation of the lattice fermion in QCD with the chiral symmetry is
one of the most fascinating problems theoretically and practically.
Although both Wilson and Kogut-Susskind (KS) fermion formulations 
have been popularly used
for the lattice QCD simulations, some disadvantages remain in these 
formulations:
In the Wilson fermion formulation the quark mass has additive quantum 
correction
and the chiral limit is reached only by the fine tuning of the mass parameter.
As a general rule we have to take continuum limit tuning the mass
appropriately in order to simulate massless QCD.
In the KS fermion formulation the number of flavors is restricted
and the original flavor symmetry is broken explicitly to some
residual one.

The domain-wall fermion formulation, which was originally proposed to define 
lattice chiral gauge theories\cite{Kaplan},
has been applied to the lattice QCD\cite{Shamir}.
This formulation is expected to have great advantage over the previous
two formulations: An advantage over the KS fermion is that the number of 
flavor is not fixed. This is manifest from its definition.
The other advantage over the Wilson fermion is that mass renormalization is
multiplicative ($m_{\rm eff} = Z_m m_{\rm tree}$). In other words, if a 
massless mode exists at the tree level it is stable against the quantum 
correction. This property is not a trivial one, but only an intuitive 
discussion on it has been given so far\cite{Shamir}.
On the other hand, the recent numerical simulation suggests that the stability
of the zero mode holds even non-perturbatively\cite{Blum-Soni}.
Therefore an analytical understanding of the domain-wall QCD is now needed.
The aim of this paper is to confirm the stability of the massless mode by the 
lattice perturbation theory and to give explicitly
the wave function renormalization of the quark field.

This paper is organized as follows.
In section 2 we will give basic tools for the perturbative calculation
with the domain-wall fermion.
It is enough to present only fermion propagator because other
Feynman rules of gauge interaction and gauge propagator are
exactly identical to that of the ordinary Wilson fermion.
In section 3 we calculate one loop corrections to the fermion
propagator.
Section 4 is the main part of this paper, where we discuss the renormalization
of the zero mode or massless quark field.
We take the diagonal basis of the mass matrix of the domain-wall fermion
and see that the zero mode is stable against the one loop correction.
The wave function renormalization factor of the massless quark field
is also given explicitly.
Section 5 is devoted to the mean field analysis.
We show that
properties of the zero mode observed
in the numerical simulation\cite{Blum-Soni}
are well explained in this approximation.
In section 6 we give our conclusion and discussion.
In appendices some derivations of formulae used in the text are presented.

In this paper we set the lattice spacing $a=1$ and
take the $SU(N_c)$ gauge group with the gauge coupling constant $g$
and the second Casimir $C_2 =\displaystyle\frac{N_c^2-1}{2N_c}$.
We set $N_c=3$ in the numerical calculations.

%
\section{Perturbation Theory with Domain-Wall Fermion} 

\subsection{Action}

We adopt the domain-wall fermion of Shamir type \cite{Shamir}
to describe massless quarks.
The domain-wall fermion is a variant of the Wilson fermion with
sufficiently many flavors and special form of the mass matrix.
Although it is also interpreted as a five dimensional Wilson fermion
\cite{Kaplan},
we prefer to treat it as the multi-flavor system\cite{NN1}.

In this point of view only difference from the Wilson fermion action
is the fermion bilinear term.
If we separate the QCD action for lattice perturbation theory
into fermion and gauge parts,
\begin{eqnarray}
S = S_{\rm fermion} + S_{\rm gauge} + S_{\rm GF} + S_{\rm FP}
 + S_{\rm measure},
\end{eqnarray}
the lattice gauge action $S_{\rm gauge}$,
the gauge fixing and the FP-ghost term $S_{\rm GF} + S_{\rm FP}$,
the invariant measure term $S_{\rm measure}$
and the gauge-fermion interaction terms in $S_{\rm fermion}$
are exactly same as those in the ordinary Wilson fermion perturbation
theory \cite{KNS,Karsten-Smit} with many flavors.

The domain-wall fermion action $S_{\rm fermion}$ is written as
\widetext
\begin{eqnarray}
S_{\rm fermion} &=& \sum_{n, m} 
\bpsi_{m,s} \left( \gamma_\mu D_{\mu} \right)_{m,n} \psi_{n,s}
+ \bpsi_{m,s} W_{m,n}^{+ \, s,t} P_{+} \psi_{n,t}
+ \bpsi_{m,s} W_{m,n}^{- \, s,t} P_{-} \psi_{n,t}
\nn\\
&+& 
m_q \bpsi_{m,s} \left( \delta_{m,n} \delta_{s,N_s} \delta_{t,1} P_{+}
+  \delta_{m,n} \delta_{s,1} \delta_{t,N_s} P_{-} \right) \psi_{n,t}
\label{eqn:fermion-action}
\end{eqnarray}
\narrowtext
where $m, n$ is four dimensional space index,
$s, t=1, \cdots, N_s$ is the flavor index.
Here the Dirac operator is given by
\begin{eqnarray}
(\gamma_\mu D_\mu)_{n, m} = \sum_{\mu}
\frac{1}{2} \gamma_\mu
\left( U_{n, \mu} \delta_{n+\hat{\mu}, m}
- U_{m, \mu}^\dagger \delta_{n-\hat{\mu}, m} \right).
\end{eqnarray}
and mass matrix $W^{\pm}_{s,t}$ is
defined as
\begin{eqnarray}
W^{\pm}_{n,m ; s,t} &=& \delta_{s \pm 1, t} \delta_{n,m} - W_{n,m} \delta_{s,t}
\end{eqnarray}
where 
\widetext
\begin{eqnarray}
W_{n, m} &=& (1-M) \delta_{n, m} 
+\frac{r}{2} \sum_\mu 
\left( U_{n, \mu} \delta_{n+\hat{\mu}, m}
+ U_{m, \mu}^\dagger \delta_{n-\hat{\mu}, m} -2 \delta_{n, m} \right)
\end{eqnarray}
\narrowtext
is a sum of the Dirac mass term and the Wilson term,
which contain gauge fields at this stage, and
$r$ is the Wilson parameter, which we set $r=-1$.
The parameter $m_q$ is the current quarks mass,
but in this paper we only treat the massless QCD taking $m_q=0$.
$P_\pm$ is a projection operator defined by
\begin{eqnarray}
P_\pm = \frac{1 \pm \gamma_5}{2}
\end{eqnarray}

In our domain-wall fermion action \eqn{eqn:fermion-action}
we have Dirac mass $M$ besides the current quark mass $m_q$.
Here we have to notice that $M$ is not the physical quark mass
but it is rather an unphysical mass of the cutoff order ($1/a$) like
Wilson term.
As will be mentioned later $M$ has an important role as a parameter of
the theory: choosing a suitable value for $M$ we have
a massless fermion mode for the vanishing current quark mass
($m_q = 0$).

In order to see the massless fermion mode it is more convenient
to be in the momentum representation
and pull out the bilinear term.
The fermion action in the momentum space is written as
\widetext
\begin{eqnarray}
S_{\rm fermion}
&=&
\int \frac{d^4 p}{(2\pi)^4}
\bpsi (-p)_s \left[ \sum_\mu i \gamma_\mu \sin p_\mu
+ W^+ (p)_{s,t} P_+ + W^- (p)_{s,t} P_- \right] \psi (p)_t
\nn\\
&+&
S_{\rm int.}
\end{eqnarray}
\narrowtext
where the mass matrix has the following form,
\begin{eqnarray}
W^{+} (p)_{s,t} &=& \delta_{s+1,t} - W(p) \delta_{s,t}
\nn\\
&=& \pmatrix{
-W(p) & 1     &        &       \cr
      & -W(p) & \ddots &       \cr
      &       & \ddots & 1     \cr
      &       &        & -W(p) \cr
}
\label{eqn:mass-matrix-p}
\\
W^{-} (p)_{s,t} &=& \delta_{s-1,t} - W(p) \delta_{s,t}
\nn\\
&=& \pmatrix{
-W(p) &        &        &       \cr
1     & -W(p)  &        &       \cr
      & \ddots & \ddots &       \cr
      &        & 1      & -W(p) \cr
}
\label{eqn:mass-matrix-m}
\\
W(p) &=& 1-M -r \sum_\mu (1-\cos p_\mu).
\end{eqnarray}
The gauge interaction term
$S_{\rm int.}$ is identical to that of
the Wilson fermion perturbation theory with $N_s$ flavors.

As will be discussed in appendix C,
in spite of the presence of the Dirac mass $M$
this fermion system has one massless fermion mode and $N_s -1$
excited modes with the mass of cut-off order 
by virtue of this mass matrix form, 
provided that $|W(p \sim 0)| < 1$ is satisfied by
a suitable choice of the Dirac mass $M$.
Here we take the momentum region $p_\mu \sim 0$ to see the zero mode
with physical momenta.
At the momentum $p_\mu \sim \pi$, where the doubler emerges in the
naive fermion formulation, the parameter condition is not satisfied
$( | W(p \sim \pi) | >1)$, so that all $N_s$ fermion modes have
mass of the cut-off order.

\subsection{Fermion Propagator}

In the next section we will calculate the one loop correction to
the fermion propagator.
In this subsection we set up the lattice Feynman rules for domain-wall
fermion with vanishing current quark mass $(m_q = 0)$.

As is discussed in the previous subsection, the domain-wall fermion
action is almost same as that of the ordinary Wilson fermion's
one with $N_s$ flavors.
The peculiar feature of the domain-wall fermion is the form of
the fermion propagator, which is given by
\begin{eqnarray}
S_{\rm F} (p)_{s,t}
=
\left[ i\gamma_\mu \bar p_\mu + W^+ (p) P_+ + W^- (p)P_- \right]^{-1}_{s,t}
\end{eqnarray}
where $\bar p_\mu = \sin (p_\mu)$.
The explicit form is written as
\widetext
\begin{eqnarray}
S_{\rm F} (p)_{s,t}
=
\left[
(-i \gamma_\mu \ovl{p}_\mu +W^-) G_R (s,t) P_{+}
+
(-i \gamma_\mu \ovl{p}_\mu +W^+) G_L (s,t) P_{-}
\right]_{s t}
\label{eqn:fermion-propagator}
\end{eqnarray}
where
\begin{eqnarray}
G_R (s, t) 
&\equiv& \left( \frac{1}{\ovl{p}^2 + W^+ W^-} \right)_{s t}
\nn\\
&=& G^0 (s-t) 
+ A_{++} e^{\alpha (s+t)}
+ A_{+-} e^{\alpha (s-t)}
+ A_{-+} e^{\alpha (-s+t)}
+ A_{--} e^{\alpha (-s-t)}
\label{eqn:GR}
\end{eqnarray}

\begin{eqnarray}
G^0 (s-t)
&=&
A \left( e^{\alpha (N_s-|s-t|)} + e^{- \alpha (N_s-|s-t|)} \right)
\\
\pmatrix{A_{++} \cr A_{-+} \cr}
&=& \frac{A}
{e^{\alpha N_s} (1-W e^\alpha) - e^{-\alpha N_s} (1-W e^{-\alpha})}
\pmatrix{
(1-W e^{-\alpha}) (e^{-2\alpha N_s} -1) \cr
W(e^{\alpha} - e^{-\alpha})           \cr
}
\\
 \pmatrix{A_{+-} \cr A_{--} \cr}
&=& \frac{A}
{e^{\alpha N_s} (1-W e^\alpha) - e^{-\alpha N_s} (1-W e^{-\alpha})}
\pmatrix{
W(e^{\alpha} - e^{-\alpha})           \cr
(1 -W e^{\alpha}) (1 - e^{2\alpha N_s}) \cr
}
\end{eqnarray}
and
\begin{eqnarray}
G_L (s, t) 
&\equiv & \left(\frac{1}{\ovl{p}^2 + W^- W^+} \right)_{s t}
\nn\\
& = & G^0 (s-t) 
+ B_{++} e^{\alpha (s+t)}
+ B_{+-} e^{\alpha (s-t)}
+ B_{-+} e^{\alpha (-s+t)}
+ B_{--} e^{\alpha (-s-t)}
\label{eqn:GL}
\end{eqnarray}
\begin{eqnarray}
\pmatrix{B_{++} \cr B_{-+} \cr}
&=& \frac{A}
{e^{\alpha N_s} (1-W e^\alpha) - e^{-\alpha N_s} (1-W e^{-\alpha})}
\pmatrix{
e^{-\alpha} (e^{-\alpha} - W) (e^{-2\alpha N_s} -1) \cr
W(e^{\alpha} - e^{-\alpha})           \cr
}
\\
\pmatrix{B_{+-} \cr B_{--} \cr}
&=& \frac{A}
{e^{\alpha N_s} (1-W e^\alpha) - e^{-\alpha N_s} (1-W e^{-\alpha})}
\pmatrix{
W(e^{\alpha} - e^{-\alpha})           \cr
e^{\alpha} (e^{\alpha} -W) (1 - e^{2\alpha N_s}) \cr
}.
\end{eqnarray}
Here
$\alpha$ and $A$ are defined as
\begin{eqnarray}
\cosh \alpha &\equiv& \frac{1 + W^2 + \ovl{p}^2}{2W(p)}
\label{eqn:alpha}
\\
\sinh \alpha &=& \frac{1}{2W}
\sqrt{(1-W^2 )^2 + 2 (1+W^2 ) \sum \sin^2 p_\mu 
+(\sum \sin^2 p_\mu)^2 }
\\
A &\equiv& \frac{1}{2 W \sinh \alpha} \frac{1}{2 \sinh (\alpha N_s)}
\end{eqnarray}
\narrowtext
Note that the argument $p$ of $W$ and $\alpha$ is suppressed
throughout this paper unless necessary.
Since this fermion propagator is invariant under
$\alpha \to -\alpha$,
we take the $\alpha > 0$ without loss of generality.
$G_R$ and $G_L$ are also symmetric in $(s,t)$.
See appendix B for the derivation.
In the one-loop calculation we use the above propagator in the
$N_s \to \infty$ limit.

\section{One Loop Calculation}

\subsection{Diagrams}

\begin{figure}[bth]
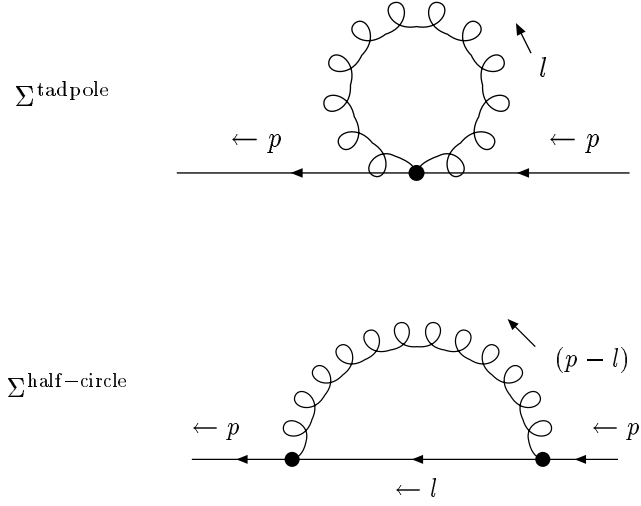

\centerline{\epsfxsize=8.5cm \epsfbox{fig1a.epsf}}
\centerline{\epsfxsize=8.5cm \epsfbox{fig1b.epsf}}
\caption{Diagrams which contribute to the one-loop correction
to the fermion propagator. Above: Tadpole diagram.
Below: Half-circle diagram.
}
\label{fig:diagram}
\end{figure}

In this section we calculate the one loop correction to
the fermion propagator, which is given by two contributions
$ \Sigma^{\rm tadpole} (p) + \Sigma^{\rm half-circle} (p)$, 
from diagrams in 
Fig.~\ref{fig:diagram}.

The 1PI fermion 2-point vertex function is given by
\widetext
\begin{eqnarray}
V^{(2)}_{\rm 1-loop} (p)_{s,t} =
\left[ i \gamma_\mu \sin p_\mu + W^+ (p) P_+ + W^- (p) P_- 
       -\Sigma (p) \right]_{s,t}
\end{eqnarray}
\narrowtext
with
\begin{eqnarray}
\Sigma (p) = \Sigma^{\rm tadpole} (p) + \Sigma^{\rm half-circle} (p) .
\end{eqnarray}

In order to investigate the massless mode of
$\Gamma^{(2)}_{\rm 1-loop} (p)_{s,t}$ in the $p_\mu \to 0$ limit,
we need only the first few terms in the $p_\mu$ expansion.
Since the only dimensionful quantity is the external momentum $p_\mu$ 
in our calculation, the higher order terms
in the $p_\mu$ expansion are also higher order in $a$.

\subsection{Contribution from Tadpole Diagram}

The contribution from the tadpole diagram is written as
\widetext
\begin{eqnarray}
\Sigma^{\rm tadpole} &=& \frac{1}{2} g^2 C_2 \sum_\mu
(i \gamma_\mu \sin p_\mu - r \cos p_\mu )
\int_{-\pi}^\pi \frac{d^4 l}{(2 \pi)^4}
\frac{1}{4 \sin^2 \frac{l}{2}} \delta_{s,t}
\\
&=&
g^2 C_2 T \left( \frac{1}{2} i \pslash + 2 \right)
\delta_{s,t} + {\cal O} (a)
\label{eqn:tadpole}
\end{eqnarray}
\narrowtext
where $T$ is the tadpole loop integral
\begin{eqnarray}
T = \int_{-\pi}^\pi \frac{d^4 l}{(2 \pi)^4}
\frac{1}{4 \sin^2 \frac{l}{2}} = 0.154612 .
\end{eqnarray}
The first term in \eq{eqn:tadpole} is finite, and the second term
linearly diverges in the limit $a \to 0$.
We see that $\Sigma^{\rm tadpole}$ is diagonal in flavor space,
and its effect is to modify the mass parameter
$M \to \wt{M}=M-2g^2 C_2 T$.

\subsection{Contribution from Half Circle Diagram}

The contribution from the half circle diagram in the Feynman gauge
\widetext
\begin{eqnarray}
\Sigma^{\rm half-circle}_{s,t} &=& \int_{-\pi}^\pi \frac{d^4 l}{(2 \pi)^4}
\sum_\mu
(-ig T^a)
\left\{ \gamma_\mu \cos \frac{1}{2} (l_\mu + p_\mu )
- ir \sin \frac{1}{2} (l_\mu + p_\mu ) \right\} 
\nn\\
&\times& S_{\rm F} (l)_{s,t}
\times (-ig T^a)
\left\{ \gamma_\mu \cos \frac{1}{2} (l_\mu + p_\mu )
- ir \sin \frac{1}{2} (l_\mu + p_\mu ) \right\} 
\times \frac{1}{(\widehat{p-l})^2}
\label{eq:half}
\end{eqnarray}
\narrowtext
cannot be calculated analytically because of its complicated 
dependence on the flavor indices $s, t$ in 
the fermion propagator.
Here $\hat{p_\mu}$ is defined as $\hat{p}_\mu=2\sin (p_\mu/2)$.

It is easily seen that the loop integral of (\ref{eq:half})
has infra-red divergence. As is in the ordinary lattice perturbation
theory the infra-red divergence can be written in an analytic form.
To do this 
we separate $\Sigma^{\rm half-circle}_{s,t}$ as follows:
\begin{eqnarray}
\Sigma^{\rm half-circle}_{s,t}(p) =
\Sigma^{\rm lat.}_{s,t}(p) + \Sigma^{\rm cont.}_{s,t}(p) 
\end{eqnarray}
where
\begin{eqnarray}
\Sigma^{\rm lat.}_{s,t}(p)  = \Sigma^{\rm half-circle}_{s,t}(p) -
\Sigma^{\rm cont.}_{s,t}(p) 
\end{eqnarray}
and $\Sigma^{\rm cont.}_{s,t}(p)$ is introduced to extract the infra-red 
divergence:
\widetext
\begin{eqnarray}
\Sigma^{\rm cont.}_{s,t}(p) =
2 g^2 C_2
\int \frac{d^4 l}{(2\pi)^4}\frac{-i\lslash( C_+P_+ + C_-P_-)_{s,t}}
{l^2 (p-l)^2}\theta(\pi^2-l^2) \nn
\end{eqnarray}
with $(C_+)_{s,t} = (1-w_0^2) w_0^{s+t-2}$,
$(C_-)_{s,t} = (1-w_0^2) w_0^{2N_s-s-t}$,
and $w_0 = W(0)$.
\narrowtext
In order to have zero modes with the physical momentum,
$w_0$ should be in the region $w_0^2 \le 1$. 
This leads to the condition of $M$ that $0\le M\le 2$.
Since $\Sigma^{\rm lat.}_{s,t}(p)$ is infra-red finite in $p\rightarrow 0$
limit, we can evaluate it in the $p$ expansion:
\begin{equation}
\Sigma^{\rm lat.}_{s,t}(p) = \Sigma^{\rm lat.}_{s,t}(0) +
p_\mu \frac{\partial \Sigma^{\rm lat.}_{s,t}}{\partial p_\mu}(0)
+O(a) .
\label{eq:expand}
\end{equation}
The logarithmically divergent part $\Sigma^{\rm cont.}(p)$ can be calculated
analytically, while
a linearly divergent and finite terms (the first and the second terms 
in eq.(\ref{eq:expand})) have to be evaluated by numerical integrations of
loop momenta. After a little algebra
we have
\widetext
\begin{eqnarray}
\Sigma^{\rm half-circle}_{s,t} =
- \left[
i \pslash \left( I^+_{s, t} P_+ + I^-_{s, t} P_- \right)
+ M^+_{s, t} P_+ + M^-_{s, t} P_-
\right]
\end{eqnarray}
where $I^\pm$ and $M^\pm$ are given by
\begin{eqnarray}
I^\pm_{s, t}
&=&
I^\pm_{\log} (s,t) + I^\pm_{\rm finite} (s,t)
\\
I_{\rm log}^\pm (s,t)
&=&
\frac{1}{16\pi^2} g^2 C_2 (C_\pm)_{s,t}
\left( \ln (\pi^2) +\frac{1}{2} - \ln p^2 \right)
\\
I_{\rm finite}^+ (s,t)
&=&
g^2 C_2 \int \frac{d^4 l}{(2\pi)^4}
\frac{1}{\hat{l}^2}
\Biggl[
\frac{1}{8}\sum_\mu\left(
\cos l_\mu ( W^-G_R + W^+G_L)(s,t) + \sin^2l_\mu (G_L+G_R)(s,t) 
\right) \nn\\
&+& \sum_\mu \frac{\sin^2 l_\mu}{4(\hat{l}^2)^2}
\left(
( W^-G_R + W^+G_L)(s,t) + 2(\sum_\nu \cos^2 l_\nu/2 -
2\cos^2 l_\mu/2)G_L(s,t) \right. \nn \\
&+& \left. \sum_\nu \sin^2 l_\nu/2 G_R(s,t)
\right) \Biggr]
-g^2 C_2 (C_+)_{s,t} \int \frac{d^4 l}{(2\pi)^4}
\frac{1}{(l^2)^2} \theta (\pi^2-l^2)
\label{eqn:finite-p}
\\
I_{\rm finite}^- (s,t)
&=&
g^2 C_2 \int \frac{d^4 l}{(2\pi)^4}
\frac{1}{\hat{l}^2}
\Biggl[
\frac{1}{8}\sum_\mu\left(
\cos l_\mu ( W^-G_R + W^+G_L)(s,t) + \sin^2l_\mu (G_L+G_R)(s,t) 
\right) \nn\\
&+& \sum_\mu \frac{\sin^2 l_\mu}{4(\hat{l}^2)^2}
\left(
( W^-G_R + W^+G_L)(s,t) + 2(\sum_\nu \cos^2 l_\nu/2 -
2\cos^2 l_\mu/2)G_R(s,t) \right. \nn \\
&+&\left. \sum_\nu \sin^2 l_\nu/2 G_L(s,t)
\right) \Biggr]
-g^2 C_2 (C_-)_{s,t} \int \frac{d^4 l}{(2\pi)^4}
\frac{1}{(l^2)^2} \theta (\pi^2-l^2)
\label{eqn:finite-m}
\\
M^+_{s,t}
&=&
g^2 C_2 \int \frac{d^4 l}{(2\pi)^4}
\frac{1}{\hat{l}^2}
\sum_\mu
\Biggl[
\cos^2 l_\mu/2 (W^+G_L)(s,t)-\sin^2 l_\mu/2(W^-G_R)(s,t)
\nn \\
&+&
\frac{1}{2}\sin^2 l_\mu ( G_L+G_R)(s,t)
\Biggr]
\label{eqn:linear-div-p}
\\
M^-_{s,t}
&=&
g^2 C_2 \int \frac{d^4 l}{(2\pi)^4}
\frac{1}{\hat{l}^2}
\sum_\mu
\Biggl[
\cos^2 l_\mu/2 (W^-G_R)(s,t)-\sin^2 l_\mu/2(W^+G_L)(s,t)
\nn \\
&+&
\frac{1}{2}\sin^2 l_\mu ( G_L+G_R)(s,t)
\Biggr]
\label{eqn:linear-div-m}
\end{eqnarray}
\narrowtext
By the dimensional counting $I^\pm$ has $\ln a^2$ divergence
and constant terms in $a$, and $M^\pm$ has $1/a$ linear divergence
when lattice spacing $a$ is introduced explicitly.
Although $M^\pm$ may have $\ln a^2$ divergence naively,
it is canceled by the algebraic relation
\begin{eqnarray}
&& ( W^+ (p=0) )_{s, t} (1-M)^t = 0
\\
&& ( W^- (p=0) )_{s, t} (1-M)^{-t} = 0.
\end{eqnarray}

The logarithmic divergence $\ln a^2$ in $I^\pm$ is given analytically.
As we can see from the form of $(C_+)_{s,t}
=(1-w_0^2) w_0^{s+t-2}$, $I^+_{\log}$ 
is localized in the boundary $(s, t)=(1, 1)$.
This is because the logarithmic divergence comes from the effect of
massless fermion mode which is localized in the boundary.
The other one $I^-_{\log}$ is also localized in the other boundary
$(s, t)=(N_s, N_s)$.

The finite terms and linearly divergent terms should be calculated
by repeating the numerical integration $O(N_s^2)$ times.
However, as can be seen in the next section,
such huge number of integrations can be avoided for the wave-function
renormalization of the quark field.
On the other-hand, since the structures of $I^\pm_{\rm finite}$ and
$M_\pm$ are useful to understand the domain-wall QCD more deeply,
we will give them in the separated paper.


\section{Renormalization of quarks field}
\label{sec:renor}

The result obtained in the previous section is summarized in the following 
form of the effective action for 2 point function with the scale 
$p^2=(\mu a)^2$ at 1-loop level:
\widetext
\begin{eqnarray}
\Gamma^{(2)} =
\bpsi (-p)_s
\left[
  i \gamma_\mu p_\mu \left( Z^+ P_+ + Z^- P_- \right)
+ \ovl{W}^+ P_+ + \ovl{W}^- P_- 
\right]_{s , t}
\psi (p)_t
\end{eqnarray}
\narrowtext
where
\begin{eqnarray}
Z^\pm &=& 1 + g^2 C_2 ( I_{\rm tad} + I^\pm_{\rm log}
       + I^\pm_{\rm finite} )
\label{eqn:z-factor}
\\
\ovl{W}^\pm &=& W^\pm(0) + g^2 C_2 ( M_{\rm tad} + M^\pm )
\end{eqnarray}
with
\begin{eqnarray}
I_{\rm tad} (s, t) &=& -\frac{1}{2} T \delta_{s, t}
= -0.077306 \delta_{s, t}
\\
M_{\rm tad}(s, t) &=& -2T \delta_{s, t} = - 0.309224 \delta_{s, t} .
\end{eqnarray}
The expressions for $I^\pm_{\log}$, $I^\pm_{\rm finite}$ and $M^\pm$
 are given in the previous section.
In this section we consider the renormalization of zero modes, which is
interpolated by the quark field:
$
q(p) = P_+ \psi(p)_1 + P_-\psi(p)_{N_s} 
$ .
Here we only present the results and give the detail of derivations
in appendix C.

\subsection{Diagonalization of mass matrix and stability of zero modes}
For the renormalization of zero modes, it is better to use new basis,
$\psi^d(p)$
which diagonalize mass matrices $\ovl{W}^{\pm}$.
This basis are given by the relation that
\begin{equation}
\psi^d_s (p) = U_{s,t}P_+\psi(p) + V_{s,t} P_-\psi(p),
\end{equation}
where unitary matrices $U$ and $V$ satisfy
\begin{eqnarray}
\left[ U \ovl{W}^-\ovl{W}^+ U^\dagger \right]_{s,t} &=& M_s^2 \delta_{s,t}
\nn \\
\left[ V \ovl{W}^+\ovl{W}^- V^\dagger \right]_{s,t} &=& M_s^2 \delta_{s,t} .
\nn
\end{eqnarray}
In our notation the mass eigen-value squared $M_s^2$ is arranged in 
such a way that $M_{N_s}^2$=0, and we can take $U$ and $V$ real matrices 
without loss of generality.

We calculate $U$ and $V$ at 1-loop level:
\begin{equation}
U = (1 + g^2 U_1)U_0 , \qquad V=(1+g^2 V_1) V_0,
\end{equation}
where tree level matrices $U_0$ and $V_0$ are analytically obtained 
in the large $N_s$ limit as follows:
\begin{equation}
[U_0]_{s,t} =
\left\{
\begin{array}{ll}
(2/N_s)^{1/2} \sin \alpha_s ( N_s + 1 - t) & s\not= N_s \\
(1-w_0^2)^{1/2} w_0^{(t-1)} & s = N_s \\
\end{array}
\right.
\end{equation}
and $
[V_0]_{s,t} = [U_0]_{s, N_s+1-t} 
$ ,
where $w_0 = 1-\wt{M}$ with $\wt{M}=M+4(u-1)$.
Here $u=1$ for the naive perturbation theory, while
$u=1-g^2 C_2 T/2$ for the tadpole improved perturbation theory.
Hereafter we will call both cases ``the tree level'' and will
not distinguish the two cases unless necessary.
If we expand the mass eigen-value squared as
$(M^2)_s =(M_0^2)_s + g^2 (M_1^2)_s$,
the tree level one is related to the phase factor $\alpha_s$ such that
$2 w_0 \cos\alpha_s = 1+w_0^2 - (M_0^2)_s$.
This phase factor, which also
satisfies $ \sin\alpha_s N_s = w_0 \sin\alpha_s (N_s + 1)$,
is explicitly given as $\alpha_s = \pi s /N_s$ in the large $N_s$ limit.
It is also shown that $U_0$ and $V_0$ diagonalize $W^\pm$ itself such that
$ [ V_0 W^+ U_0^\dagger ]_{s,t}= [ U_0 W^- V_0^\dagger ]_{s,t}= 
(M_0)_s\delta_{s,t}$.

We now consider $\ovl{W}^\pm$, which is denoted
as $\ovl{W}^\pm = W_0^\pm + g^2 W_1^\pm$, where
$(W_0^\pm)_{s,t} = W^\pm (0) = \delta_{t,s\pm 1}-w_0$ 
and $g^2(W_1^\pm)_{s,t} = g^2C_2(M^\pm+M_{\rm tad})_{s,t}
+4(1-u)\delta_{s,t}$.
To diagonalize $\ovl{W}^\mp\cdot\ovl{W}^\pm$ at 1-loop order,
$U_1$ and $V_1$ should satisfy
\widetext
\begin{eqnarray}
(U_1)_{s,t} (M_0^2)_t &+& (M_0^2)_s (U_1^\dagger)_{s,t} \nn \\
&+&
(U_0 W_1^- V_0^\dagger\cdot V_0 W_0^+ U_0^\dagger)_{s,t}
+(U_0 W_0^- V_0^\dagger\cdot V_0 W_1^+ U_0^\dagger)_{s,t} 
= (M_1^2)_s\delta_{s,t} , \\
(V_1)_{s,t} (M_0^2)_t &+& (M_0^2)_s (V_1^\dagger)_{s,t} \nn \\
&+&
(V_0 W_1^+ U_0^\dagger\cdot U_0 W_0^- V_0^\dagger)_{s,t}
+(V_0 W_0^+ U_0^\dagger\cdot U_0 W_1^- V_0^\dagger)_{s,t} 
= (M_1^2)_s\delta_{s,t} .
\end{eqnarray}
\narrowtext
Using the fact that $(U_1,V_1)_{s,t} = - (U_1,V_1)_{t,s}$ implied by
the unitarity and the reality, and  $V_0 W_0^+ U_0^\dagger$ and 
$U_0 W_0^- V_0^\dagger$ are diagonal, we can easily solve the above equation
as
\begin{eqnarray}
(U_1)_{s,t} &=& \frac{ (M_0)_t (\wt{W_1})_{t,s}+ (M_0)_s(\wt{W_1})_{s,t}}
{(M_0^2)_s - (M_0^2)_t} \nn \\
(V_1)_{s,t} &=& \frac{ (\wt{W_1})_{s,t}(M_0)_t+ (\wt{W_1})_{t,s}(M_0)_s}
{(M_0^2)_s - (M_0^2)_t} \nn 
\end{eqnarray}
for $s\not= t$, and
\begin{equation}
(M_1^2)_s = 2 (\wt{W_1})_{s,s} (M_0)_s, \qquad (U_1)_{s,s}=(V_1)_{s,s}= 0,
\end{equation}
where $\wt{W_1} = V_0 W_1 U_0^\dagger $.
The mass eigen-value squared $M_s^2=(M_0^2)_s+g^2 (M_1^2)_s$
obtained above leads to the mass eigen-value $M_s$ itself:
$M_s = (M_0)_s + g^2 (\wt{W_1})_{s,s}$.
Note that $M_{N_s} = 0$ since $(M_0)_{N_s}=0$ and 
$(\wt{W_1})_{N_s,N_s}=0$ in the large $N_s$ limit
as is shown in appendix C.
This result explicitly demonstrates the stability of the zero modes 
against 1-loop corrections in domain-wall QCD.
As in the case at the tree level, it is shown that
\begin{equation}
(V \ovl{W}^+ U^\dagger)_{s,t} = (U \ovl{W}^-V^\dagger)_{s,t} = 
M_s \delta_{s,t} + O(g^4) .
\end{equation}

\subsection{Wave function renormalization for quark fields}
After diagonalization of the mass matrix, the effective action
for the zero mode field $\psi^d(p)_{N_s} = \chi_0(p)$ becomes
\begin{equation}
\bar\chi_0(-p) \left[i \gamma_\mu p_\mu \left( \wt{Z_+} P_+ + \wt{Z_-} 
P_-\right)
\right] \chi_0(p)
\end{equation}
where
\widetext
\begin{equation}
\wt{Z_\pm} 
= 1-g^2C_2\frac{T}{2} 
+ \frac{g^2C_2}{16\pi^2} \left( \log \pi^2 + \frac{1}{2} - \log(\mu a)^2 
\right)
+g^2 (I^d_\pm)_{N_s,N_s}
\end{equation}
with
$ I^d_+ = C_2  (U_0 I_{\rm finite}^+ U_0^\dagger) $ and
$ I^d_- = C_2  (V_0 I_{\rm finite}^- V_0^\dagger) $ .
Since the interpolating quark field $q(p)$ is expressed
as
$ q (p)=( U_{N_s1}P_+ + V_{N_s,N_s} P_-) \chi_0 (p)$, and
$
\langle \chi_0(p) \bar \chi_0(-p) \rangle =
\left[\displaystyle\frac{1}{\wt{Z_+}}P_+ 
+ \displaystyle\frac{1}{\wt{Z_-}}P_-\right]
\displaystyle\frac{-i\gamma_\mu p_\mu}{p^2}
$ ,
we obtain
\narrowtext
\begin{equation}
\langle q(p) \bar q(-p) \rangle =
\left[\frac{U_{N_s1}^2}{\wt{Z_+}}P_+ + \frac{V_{N_s,N_s}^2}{\wt{Z_-}}P_-\right]
\frac{-i\gamma_\mu p_\mu}{p^2} .
\end{equation}
Therefore, the renormalized quark field $Q(p)$, which satisfies
$\langle Q(p)\bar Q(-p)\rangle =\displaystyle  \frac{-i\gamma_\mu p_\mu}{p^2}$,
is given by $Q(p) = [ (Z_F^+)^{1/2}P_+ + (Z_F^-)^{1/2}P_- ] q(p)$ with
$ Z_F^+ = \displaystyle \frac{\wt{Z_+}}{U_{N_s1}^2}$ and
$ Z_F^- = \displaystyle \frac{\wt{Z_-}}{V_{N_s,N_s}^2}$.
Since an explicit evaluation shows that
$(I^d_+)_{N_s,N_s} = (I^d_-)_{N_s,N_s} \equiv I^d $, thus
$\wt{Z_+} =\wt{Z_-} \equiv \wt{Z}$, and
$ (U_{N_s1})^2 = (V_{N_s,N_s})^2 = 1-w_0^2 $ ,
we finally obtain $Z_F^+ = Z_F^- \equiv Z_F =\displaystyle 
\frac{\wt{Z}}{1-w_0^2}$
where
\widetext
\begin{equation}
\wt{Z} = 1-g^2C_2\frac{T}{2} +\displaystyle
\frac{g^2}{16\pi^2} C_2 ( \log \pi^2 + \frac{1}{2} - \log(\mu a)^2 )
+g^2 I^d .
\end{equation}
Here one unknown constant $I^d$ is given by
\begin{eqnarray}
I^d & = & C_2 \int\frac{d^4l}{(2\pi)^4}\left\{
\frac{1}{8\hat l^2} \sum_\mu \left[  \sin^2l_\mu ( \wt{G_R}+\wt{G_L})
+ 2 \cos l_\mu (w_0-W(l))\wt{G_R} \right] \right. \nn \\
&+& \sum_\mu \frac{\sin^2 l_\mu}{2(\hat l^2)^2} 
\left[(w_0-W(l))\wt{G_R}+(\sum_\nu \cos^2 l_\nu/2 - 2\cos^2l_\mu/2 )
\wt{G_L}+\sum_\nu (\sin^2l_\nu/2) \wt{G_R} \right] \nn \\
&-&\left. \frac{1}{(l^2)^2}\theta(\pi^2-l^2)\right\}
\end{eqnarray}
\narrowtext
where
\begin{eqnarray}
\wt{G_L} &=& A \left[ \wt{G} -\frac{e^{\alpha}-W}{e^{-\alpha}-W}
\frac{1}{(e^\alpha-w_0)^2} \right]\nn \\
\wt{G_R} &=& A \left[ \wt{G} -\frac{1}{(e^\alpha-w_0)^2}\right] \nn 
\end{eqnarray}
with 
\begin{eqnarray}
A &=& \frac{1-w_0^2}{2W\sinh\alpha} \nn \\
\wt{G} &=& \frac{\sinh\alpha_0 -\sinh\alpha}{2w_0\sinh\alpha_0
(\cosh\alpha_0 -\cosh\alpha )} \nn
\end{eqnarray}
and $e^{-\alpha_0}=w_0$. The numerical value of $I^d$ is given in 
Table~\ref{tab:wave} at several values of $ \wt{M}$,
together with the total 1-loop renormalization factor $Z_1$ (
$\wt{Z} \equiv 1 + g^2 Z_1$ ) at $\mu a =1$ and the ratio of the non-tadpole
contribution $(Z_1)_{\rm non-tad} \equiv I^d+
\displaystyle\frac{C_2}{16\pi^2}(\log\pi^2+0.5) = I^d + 0.02355$
to the total one. Note also that the tadpole contribution gives
$Z_{\rm tad}\equiv -C_2 T/2 = -0.1031$.
  From this table, we see that $I^d$ is small and 
depends on $\wt{M}$ very weakly:
The value $I^d = -0.01945$ at $\wt{M}=0.05$ monotonically increases
(; decreases in the absolute value )
to $I^d = -0.01222$ at $\wt{M}=0.95$.
Furthermore the non-tadpole contribution $Z_{\rm non-tad}$
is relatively small: 4\% at $\wt{M}=0.05$ and 12\% at $\wt{M}=0.95$,
so that the tadpole contribution becomes dominant
at all $\wt{M}$.
This justifies the use of the tree-level result with the tad-pole improvement.
Since $Z_1 \simeq 0.1$, the one-loop correction to the $Z$ factor is
about 10 \% at $g^2 \sim 1.0$.

\section{Mean field analysis at finite $N_s$}
\label{sec:MF}

As seen in the previous sections,
due to the presence of off-diagonal terms in the extra
dimension, analysis of the 1-loop correction to
domain-wall quarks becomes too complicated to be easily applied to
results of the numerical simulations, which should be performed
on finite $N_s$.
In this section we adopt an approximated but simpler method
to analyze the effect of 1-loop corrections.
We call the method the mean field (MF) analysis since
the link variable $U_{n,\mu}$ in the fermion action
is simply replaced by the mean field $u$ which is independent
on $n$ and $\mu$. 
After this replacement the fermion propagator
can be explicitly calculated and result is identical to the one
given in Appendix B with
the replacement such that $x\rightarrow u x$ and $\cos p_\mu
\rightarrow u\cos p_\mu$.
In perturbation theory this is equivalent to the tree level analysis
with the tad-pole improvement, which has been shown in the previous section
to give about 90 \% of the wave function renormalization factor at 1-loop 
level.

Since we are interested in the zero mode at $s=1$, we set $s=t=1$
in the propagator. In this case the zero mode appears in 
$B_{--} e^{-2\alpha}$
of $G_L$, which is given at non-zero $m_q$ by
\begin{equation}
B_{--} e^{-2\alpha}
=\frac{(1-We^{-\alpha})(1-m_q^2)}{2W\sinh(\alpha) F}
\end{equation}
where
\widetext
\begin{eqnarray}
F&=&We^\alpha-1+m_q^2(1-We^{-\alpha})-4m_q\cdot W\cdot\sinh(\alpha)
e^{-\alpha N_s} \nn \\
&+&e^{-2\alpha N_s}(1-We^{-\alpha}+m_q^2(We^\alpha-1)).
\end{eqnarray}
In the small momentum limit, this leads to
\begin{equation}
\lim_{p^2\rightarrow 0} B_{--} e^{-2\alpha}
=\frac{Z^{-1}}{p^2+ m_F^2} \times
\frac{(1-w_0^2)^2+p^2 u w_0^2}{(1-w_0^2)^2+p^2 u (1+w_0^2)}
\end{equation}
where
$
Z^{-1} =\dfrac{1-m_q^2}{A u}
$,
$
m_F^2 = \dfrac{B}{A u}
$
and $w_0=1-M+4(1-u)=1-\wt{M}$ with
\begin{eqnarray}
A&=&\frac{1}{1-w_0^2}\biggl[
1+m_q^2w_0^2-w_0(1-w_0^2)m_q^2
+m_q w_0^{N_s} \nn \\
&\times& \{ 2N_s(1-w_0^2)-1-w_0^2+2w_0(1-w_0^2)-N_s(1-w_0^2)^2/w_0\}
\nn \\
&+& w_0^{2N_s}\{ w_0^2+m_q^2-2N_s(1-w_0^2)-w_0(1-w_0^2)+N_s(1-w_0^2)^2/w_0\}
\biggr] \nn \\
B &=& (1-w_0^2)\left[m_q^2 - 2m_q w_0^{N_s}+w_0^{2N_s}\right] .
\nn
\end{eqnarray}
\narrowtext
Since the pole in the second factor is in general
larger than the physical pole in the first factor,
we neglect the second factor in the latter analysis.

\begin{figure}[bth]
\centerline{\epsfxsize=8.5cm \epsfbox{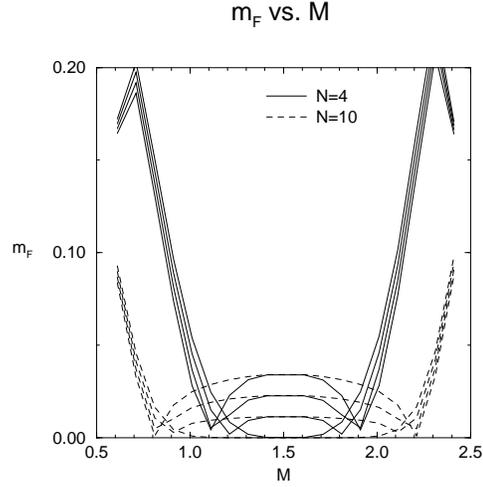}}
\caption{The fermion mass $m_F$ obtained in the mean-field approximation
as a function of $M$ for $N_s$=4 (solid lines) and $N_s$=10 (dashed line),
at $m_q$= 0, 0.01, 0.02, 0.03 from below to above around $M=1.5$.
}
\label{fig:MF}
\end{figure}

Now we use the above formula to understand the behavior of the zero
mode observed in ref.~\cite{Blum-Soni}.
For the value of $u$ there are several choices.
The tadpole diagram alone gives
\[
u = 1-g^2 C_2 T/2 = 1-0.10307 g^2 \simeq \exp[-0.10307 g^2] ,
\]
where we may take the bare coupling $2N_c/\beta$ or
the renormalized coupling $g^2_{\ovl{MS}}(\pi/a)$
for $g^2$ in the above formula.
Alternatively we may also use the ``observed'' value of $u$:
$u = P^{1/4}$ where $P$ is the average value of the plaquette
normalized to unity.
We adopt the latter one in our analysis.
The configurations in ref.~\cite{Blum-Soni}
generated at $\beta = 5.7$ and $m_q a = 0.01$ by the dynamical
Kogut-Susskind quark action
give $P = 0.5772$, which leads to $u=0.872$.
In ref.~\cite{Blum-Soni} two remarkable features are found for the zero mode:
no zero mode is observed for $N_s = 4$ and
the zero mode is observed at $M=1.7$ but not at $M \le 1.0$
for $N_s = 10$.
To explain these we calculate $m_F$ as a function of $M$
for both $N_s = 4$ and 10 at $m_q$ = 0, 0.01, 0.02 ,0.03,
and plot the results in Fig.\ref{fig:MF}, where
solid lines are for $N_s = 4$ and dashed lines for $N_s = 10$.
Four lines for each $N_s$ correspond to $m_q =$0, 0.01, 0.02, 0.03
from below to above around $M=1.5$.
The result tells us the followings.
The allowed range for the light fermion is very narrow 
for $N_s =4$ (roughly $1.4 < M < 1.6$ ).
This may be a reason why the light state could not be found
in the simulation\cite{Blum-Soni}.
Note that the allowed range for the zero mode is
$0.512 < M < 2.512$ in the $N_s \to \infty$ limit.
Although the allowed range becomes larger for
$N_s = 10$ ($ 1.1 < M < 1.9 $),
no light state appears at $M \le 1.0$, as observed in the 
simulation. Furthermore the order of the fermion mass 
$m_F$ is reversed to the order of the current quark mass $m_q$
at $M \le 1.0$:  $m_F$ is largest at $m_q = 0$.
The plot also supports the fact that the zero mode is
observed at $M=1.7$ in the simulation.

As seen in the above the behavior of the numerical simulation
is understandable by the MF analysis, which can supply
useful informations on the tuning of parameters in numerical 
simulations such as $N_s$, $M$ or $m_q$ before-hand.
For example we may take $N_s = 4$ for the simulations,
which reduces the cost of both CPU time and memory a lot,
if $M$ is appropriately chosen ( $M \simeq 1.5 $ for $U=0.872$ ).


\section{Conclusion and Discussion}
In this paper we calculate one-loop correction to the fermion propagator
in the massless lattice QCD formulated via domain-wall fermions.
We show that the zero mode is stable against the one-loop correction:
no additative counter term to the quark mass is generated in the large $N_s$
limit. This property is very different from and superior to the ordinary
Wilson fermion formulation.
We explicitly calculate the wave-function renormalization factor for the
massless quarks and show that the tadpole contribution becomes
dominant at all $\wt{M}$.
We also adopt the mean-field analysis to this model, demonstrating
that it can qualitatively explain data obtained in the numerical 
simulation\cite{Blum-Soni}.

Although our results strongly indicate that the domain-wall QCD can avoid
the fine tuning problem of the quark mass,
the mechanism which gives the zero mode in this formulation has not
been fully understood yet. Since our proof for the stability of the zero mode
contains an explicit calculation at 1-loop (: $(\wt{W_1})_{Ns,N_s}=0$),
it can not be easily carried over to higher orders.
The result of numerical simulation\cite{Blum-Soni} suggests that
the zero mode is also stable against the non-perturbative dynamics.
There may be a yet unknown symmetry which ensures the existence of
zero mode in the large $N_s$ limit. To find such a symmetry will be
important for our understanding of the formulation

In this paper only the wave-function renormalization factor is explicitly
evaluated. Based on the method developed in this paper,
it is possible to calculate more complicated quantities
such as renormalization factors for the quark mass, currents and
4-fermi operators, which are necessary to get the continuum physics 
from numerical simulations. The results of this paper also suggest
that the smeared quark operator $q^{\rm smear} =\sum_s (w_0^s P_+\psi_s
+w_0^{N_s-s} P_-\psi_s)$  may give better signals 
than $q=P_+\psi_1+P_-\psi_{N_s}$ does, since it has a larger overlap
to zero modes.

\vskip 0.7cm

After this work has been completed, there appears a new paper\cite{Neuberger},
in which the stability of the zero mode is generally considered.

\section*{Acknowledgments}
We would like to thank Dr. Izubuchi for his valuable comments and
discussion.
Discussions with Drs. Zenkin, Nagai, Kaneda, and Ishizuka 
are also helpful and encouraging.
This work is supported in part by the Grand-in Aid for
Scientific Research (Nos. 08640350, 09246206, 2373)
from the Ministry of Education, Science and Culture. 
Y.T. is a JSPS fellow.

\section*{Appendix A. Action and Feynman Rules}
\label{app:action}

The gauge part of the action is exactly same as that of the ordinary
lattice QCD action \cite{Karsten-Smit}.
\widetext
\begin{eqnarray}
&&
S_{\rm gauge} = \sum_n \sum_{(\mu \nu)}
- \frac{\beta}{N_s} {\rm Re} \; \tr
\left( U_{n, \mu}^\dagger U_{n+\hat{\nu}, \mu}^\dagger
U_{n+\hat{\mu}, \nu} U_{n, \mu} \right)
\\
&& 
S_{\rm GF} = \sum_n \frac{1}{2\alpha}
\left( \nabla_\mu A_\mu^a (n+\frac{1}{2}\hat{\mu}) \right)^2
\\
&& 
S_{\rm FP} = \sum_{n, \mu} (\bc_{n+\hat{\mu}}^a - \bc_n^a)
\left[ c_{n+\hat{\mu}}^b 
  E_{ba}^{-1} \left(g A_\mu(n+\frac{1}{2}\hat{\mu})\right)
- E_{ab}^{-1} \left(g A_\mu(n+\frac{1}{2}\hat{\mu})\right) c_n^b
\right]
\\
&&
S_{\rm measure} = -\frac{1}{2} \sum_n \sum_\mu \tr \ln
\left(
\frac{1-\cos \left( g A_\mu^c (n+\frac{1}{2}\hat{\mu}) {\rm ad}(T^c) \right)}
{ \left( g A_\mu^c (n+\frac{1}{2}\hat{\mu}) {\rm ad}(T^c) \right)^2}
\right)_{ab}
\end{eqnarray}
\narrowtext
where
$g$ is the coupling of the $SU(N_c)$ gauge,
$\beta=2N_c/g^2$.
$\alpha$ is the gauge parameter.
The actions
$S_{\rm FP}$ and $S_{\rm measure}$ is not needed in our calculation at
one loop level.

The momentum representation of gauge part is
\widetext
\begin{eqnarray}
S_{\rm gauge} + S_{\rm GF}
&=&
\frac{1}{2} \int \frac{d^4 p}{(2\pi)^4}
A_\mu^a (-p) \left[
\hat{p}^2 \delta_{\mu \nu} - (1-\frac{1}{\alpha}) \hat{p}_\mu \hat{p}_\nu
\right] A_\nu^a (p)
\nn\\
&+&
\cdots
\end{eqnarray}
\narrowtext
here $+\cdots$ denotes the gluon self interactions which do not come into
play in our calculation.

The fermion-gauge interaction terms in the momentum representation is
\widetext
\begin{eqnarray}
S_{\rm int.}
&=&
\sum_{n=1}^\infty
\int \frac{d^4 k}{(2\pi)^4} \frac{d^4 p}{(2\pi)^4}
\frac{d^4 l_1}{(2\pi)^4} \cdots \frac{d^4 l_n}{(2\pi)^4} 
(2\pi)^4 \delta^4 (k+p+l_1 + \cdots +l_n)
\nn\\
&&
\times
\frac{i^n}{n!} g^n A_\mu^{a_1} (l_1) \cdots A_\mu^{a_n} (l_n)
\bpsi (k)_s T^{a_1} \cdots T^{a_n}
\nn\\
&&
\times
\left[
\frac{\gamma_\mu}{2}
\left(
e^{\frac{i}{2} (p_\mu - k_\mu)} -(-)^n e^{-\frac{i}{2} (p_\mu - k_\mu)}
\right)
-\frac{r}{2}
\left(
e^{\frac{i}{2} (p_\mu - k_\mu)} +(-)^n e^{-\frac{i}{2} (p_\mu - k_\mu)}
\right)
\right]
\psi (p)_s.
\label{eqn:fermion-action-p}
\end{eqnarray}
\narrowtext
The domain-wall fermion propagator is already given by the
\eq{eqn:fermion-propagator}.

The fermion gluon interaction vertices are given by 
\eqn{eqn:fermion-action-p}.
Although there are infinite number of interactions in the lattice
perturbation theory, only two of them are needed for the present
purpose.
One of them is the fermion interaction vertex with one gluon field,
which is given by
\widetext
\begin{eqnarray}
V_1 (k,p;l,a;\mu)
= -i g T^a \{ \gamma_\mu \cos \frac{1}{2}(-k_\mu + p_\mu)
              -i r \sin \frac{1}{2}(-k_\mu + p_\mu) \}.
\end{eqnarray}
The other is the vertex with two gluon fields, given by
\begin{eqnarray}
V_2 (k,p;l_1,a,l_2,b;\mu)
= \frac{1}{2} g^2 \frac{1}{2} \{T^{a}, T^{b}\}
\{ i \gamma_\mu \sin \frac{1}{2}(-k_\mu + p_\mu)
-r \cos \frac{1}{2} (-k_\mu + p_\mu) \}\delta_{\mu\nu} .
\end{eqnarray}

\narrowtext
The gluon propagator is given by
\begin{eqnarray}
G_{\mu \nu}^{ab} (p)
=\frac{1}{\hat{p}^2}
\left[\delta_{\mu \nu}
- (1-\alpha) \frac{\hat{p}_\mu \hat{p}_\nu}{\hat{p}^2}
\right]
 \delta_{ab}.
\end{eqnarray}
We set $\alpha = 1$ in this paper.

\section*{Appendix B. Derivation of free fermion propagator}
\label{app:prop}

In this appendix we derive the free fermion propagator, used
in the text. 
For the later use in perturbative analyses
of this model, non-zero current quark mass $m_q$ for finite $N_s$
is considered. See also Refs.\cite{vranas,NN1,AH}. 
We also derive the propagator with
Majorana mass terms, which becomes important for
the lattice definition of the $N=1$ supersymmetric model
via domain-wall fermions\cite{Nishimura,ANZ}.

\subsection{Propagator with non-zero $m_q$}
The free fermion propagator has the following form:
\widetext
\begin{eqnarray}
S_{\rm F} (p)_{s,t}
=
\left[
(-i \gamma_\mu \ovl{p}_\mu +W_m^-) G_R (s,t) P_{+}
+
(-i \gamma_\mu \ovl{p}_\mu +W_m^+) G_L (s,t) P_{-}
\right]_{s t}
\nn
\end{eqnarray}
where
\begin{eqnarray}
G_R (s, t) 
\equiv \left( \frac{1}{\ovl{p}^2 + W_m^+ W_m^-} \right)_{s t}
\qquad {\rm and}\qquad
G_L (s, t) 
\equiv \left(\frac{1}{\ovl{p}^2 + W_m^- W_m^+} \right)_{s t} 
\nn
\end{eqnarray}
with
\begin{eqnarray}
(W_m^+)_{s,t} = (W^+)_{s,t} + m_q \delta_{s,N_s}\delta_{t,1}
\qquad {\rm and}\qquad
(W_m^-)_{s,t} = (W^-)_{s,t} + m_q \delta_{s,1}\delta_{t,N_s}
\end{eqnarray}

We first consider $G_R$.
The following equation is satisfied for $G_R$:
\begin{equation}
\sum_t \left[ (x+ W^+ W^-)_{s,t} + m_q ( W^+_{s1}\delta_{tN_s}
+ \delta_{s,N_s}W^-_{1t} ) + m_q^2\delta_{s,N_s}\delta_{tN_s}
\right] G_R(t,u) = \delta_{su}
\end{equation}
with $x =\ovl{p}^2$.
Therefore, except $s=N_s$ or 1, this equation is
satisfied by
\begin{eqnarray}
G_R(s,t)=G(s,t) + A_{++} e^{\alpha (s+t)}+ A_{+-} e^{\alpha (s-t)}
+ A_{-+} e^{\alpha (-s+t)}+ A_{--} e^{\alpha (-s-t)}
\label{eq:formGR}
\end{eqnarray}
\narrowtext
where 
\begin{eqnarray}
G(s,t) = A \left( e^{\alpha (N_s-|s-t|)} 
+ e^{- \alpha (N_s-|s-t|)} \right)
\end{eqnarray}
becomes a special solution to the equation
$ (x+W^+W^-)G_R = 1$, with
\begin{eqnarray}
&&
\cosh \alpha \equiv \frac{1 + W^2 + x}{2W(p)}
\nn\\
&& A \equiv \frac{1}{2 W \sinh \alpha} 
\frac{1}{2 \sinh (\alpha N_s)} ,
\end{eqnarray}
and other terms are general solutions to the equation
$ (x+W^+W^-)G_R = 0$. We can fix their coefficients $A_{\pm\pm}$
by a boundary condition at $s=1$:
\widetext
\begin{equation}
(x + W^2 +1)G_R(1,t)-W\cdot G_R(2,t)-W\cdot m_q\cdot G_R(N_s,t)=
\delta_{1t},
\end{equation}
\narrowtext
which is simplified to
\begin{equation}
G_R(0,t)-m_qG_R(N_s,t) = 0,
\label{eq:bcR1}
\end{equation}
and another boundary condition at $s=N_s$:
\widetext
\begin{equation}
(x + W^2 )G_R(N_s,t)-W\cdot G_R(N_s-1,t)
-W\cdot m_q\cdot G_R(1,t)+m_q^2G_R(N_s,t)=
\delta_{N_s,t},
\end{equation}
which is reduced to
\begin{equation}
G_R(N_s,t)-W\cdot G_R(N_s+1,t)+W\cdot m_q G_R(1,t)
-m_q^2 G_R(N_s,t) = 0 .
\label{eq:bcR2}
\end{equation}
Plugging eq.(\ref{eq:formGR}) into eqs.(\ref{eq:bcR1}) and 
(\ref{eq:bcR2}) leads to
\begin{eqnarray}
& &
\left(
\begin{array}{ll}
1-m_q e^{\alpha N_s} & 1-m_q e^{-\alpha N_s} \\
e^{\alpha N_s}(1-We^\alpha-m_q^2+W m_q e^{\alpha (1-N_s)}) &
e^{-\alpha N_s}(1-We^{-\alpha}-m_q^2+W m_q e^{\alpha (N_s-1)})
\end{array}
\right)
\left(
\begin{array}{ll}
A_{++} & A_{+-} \\
A_{-+} & A_--
\end{array}
\right)
\nn \\
& &
= - A
\left(
\begin{array}{ll}
e^{-\alpha N_s}-m_q &  e^{\alpha N_s}-m_q \\
1-We^{-\alpha}-m_q^2+W m_q e^{-\alpha (N_s+1)} &
(1-We^{\alpha}-m_q^2+W m_q e^{\alpha (N_s+1)})
\end{array}
\right) .
\end{eqnarray}
Solving this we obtain
\begin{eqnarray}
\left(
\begin{array}{l}
A_{++} \\
A_{-+}
\end{array}
\right) 
&=& \frac{A}{F}
\left(
\begin{array}{l}
(e^{-2\alpha N_s}-1)(1-We^{-\alpha})(1-m_q^2) \\
2W \sinh(\alpha) (1-2m_q\cosh(\alpha N_s)+m_q^2)
\end{array}
\right) \nn \\
\left(
\begin{array}{l}
A_{+-} \\
A_{--}
\end{array}
\right) 
&=& \frac{A}{F}
\left(
\begin{array}{l}
2W \sinh(\alpha) (1-2m_q\cosh(\alpha N_s)+m_q^2) \\
(1-e^{2\alpha N_s})(1-We^{\alpha})(1-m_q^2) 
\end{array}
\right) \nn
\end{eqnarray}
where
\begin{equation}
F = e^{\alpha N_s}[ 1-We^\alpha  +m_q^2 (We^{-\alpha}-1)]
+4Wm_q\sinh(\alpha)
+e^{-\alpha N_s}[ We^{-\alpha}-1 +m_q^2 (1-We^{\alpha})] .
\nn
\end{equation}

Similarly, plugging the general solution for $G_L$
\begin{eqnarray}
G_L(s,t)=G(s,t) + B_{++} e^{\alpha (s+t)}+ B_{+-} e^{\alpha (s-t)}
+ B_{-+} e^{\alpha (-s+t)}+ B_{--} e^{\alpha (-s-t)}
\label{eq:formGL}
\end{eqnarray}
into the boundary conditions
\begin{eqnarray}
G_L(N_s+1,t)-m_qG_L(1,t) &=& 0 \\
\label{eq:bcL1}
G_L(1,t)-W\cdot G_L(0,t)+W\cdot m_q G_L(N_s,t)
-m_q^2 G_L(1,t) & = & 0 ,
\label{eq:bcL2}
\end{eqnarray}
we finally obtain
\begin{eqnarray}
\left(
\begin{array}{l}
B_{++} \\
B_{-+}  
\end{array}
\right) 
&=& \frac{A}{F}
\left(
\begin{array}{l}
(e^{-2\alpha N_s}-1)e^{-\alpha}(e^{-\alpha}-W)(1-m_q^2) \\
2W \sinh(\alpha) (1-2m_q\cosh(\alpha N_s)+m_q^2)
\end{array}
\right) \nn \\
\left(
\begin{array}{l}
B_{+-} \\
B_{--}
\end{array}
\right) 
&=& \frac{A}{F}
\left(
\begin{array}{l}
2W \sinh(\alpha) (1-2m_q\cosh(\alpha N_s)+m_q^2) \\
(1-e^{2\alpha N_s})e^\alpha (e^{\alpha}-W)(1-m_q^2) 
\end{array}
\right) . \nn
\end{eqnarray}
\narrowtext

\subsection{Propagator with the Majorana mass term at $N_s$}
For an application of the free fermion propagator obtained
in the domain-wall model, we consider the model with the Majorana
mass term on the anti-boundary at $s=N_s$, which
has been proposed for a lattice definition of the $N=1$
super Yang-Mills theory\cite{Nishimura,ANZ}. Here we set $m_q =0$.

A free fermion action of the model with the Majorana mass $m_0$
can be written in the momentum space as 
\begin{equation}
S=\frac{1}{2}\bar\Psi (-p)_s D_{s,t}(p) \Psi(p)
\end{equation}
where
\begin{eqnarray}
\Psi_s(p) = \left(
\begin{array}{ll}
\psi_s(p) ,
\bar\psi_s(p)
\end{array}
\right),
\qquad
\bar\Psi_s(p) = \left(
\begin{array}{l}
\bar\psi_s(p) \\
\psi_s(p)
\end{array}
\right),
\end{eqnarray}
and
\widetext
\begin{equation}
D(p) = T_0(p) + m_0 X =
\left(
\begin{array}{cc}
D_0(p) & 0 \\
0      & -D_0(-p)^T
\end{array}
\right)
+ m_0 \delta^2 P_+ I P_-
\end{equation}
with
$(\delta^2)_{s,t}\equiv \delta_{s,N_s}\delta_{N_s,t}$,
and 
\[
P_+ = 
\left(
\begin{array}{cc}
P_+ & 0 \\
0 & P_-
\end{array}
\right), \quad
P_- = 
\left(
\begin{array}{cc}
P_- & 0 \\
0 & P_+
\end{array}
\right) , \qquad
I = 
\left(
\begin{array}{cc}
0 & I_2 \\
I_2 & 0
\end{array}
\right) 
=
\left(
\begin{array}{cccc}
0 & 0 & \sigma_2 & 0\\
0 & 0 & 0 & \sigma_2 \\
\sigma_2 & 0 & 0 & 0 \\
0 & \sigma_2 & 0 & 0
\end{array}
\right) 
\]
\narrowtext
in terms of $8\times 8$ matrices.
Here
\begin{equation}
D_0(p)=i\gamma_\mu\ovl{p}_\mu + W^+ P_+ + W^-P_- ,
\end{equation}
is an inverse of the massless free fermion propagator in 
the domain-wall QCD.

By expanding $D^{-1}$ in $m_0$ and rearranging it
we obtain
\begin{eqnarray}
D^{-1} & = &\sum_{n=0}^\infty (-T_0^{-1} m_0 P_+ I P_-\delta^2)^n
T_0^{-1} \nn \\
&=&\sum_{n=0}^\infty (-m_0)^n T_0^{-1}\delta P_+ Z^{n-1}
I P_-\delta T_0^{-1}
\end{eqnarray}
where $Z = IP_-\delta T_0^{-1} \delta P_+$.

Using $Z^2 = - x (G_R(p)_{N_s,N_s})^2 P_+$ and
suming over $n$,
we finally get
\widetext
\begin{eqnarray}
D^{-1}  = T_0^{-1} +
\left[ -m_0 T_0^{-1}\delta P_+ I P_- \delta T_0^{-1}
+m_0^2 T_0^{-1}\delta P_+ Z I P_- \delta T_0^{-1}
\right]
\times \frac{1}{1+m_0^2 x (G_R(p)_{N_s,N_s})^2} .
\end{eqnarray}

Explicitly this formula gives, in terms of $2\times 2$ block notations,
\begin{eqnarray}
D(p)^{-1}_{11} &=& -D(-p)^{-1}_{22} =
\langle \psi(p)\bar\psi(-p)\rangle \nn \\
&=&
(-i\gamma_\mu\ovl{p}_\mu)[Z_+(p)P_+ + Z_-(p)P_- ]
+ M_+(p)P_+ + M_-P_-
\end{eqnarray}
where
\begin{eqnarray}
Z_+(p)_{s,t} &=& G_R(p)_{s,t}-\frac{m_0^2 x G}{1+m_0^2 x G^2}
G_R(p)_{s,N_s}G_R(p)_{N_s,t} \nn \\
Z_-(p)_{s,t} &=& G_L(p)_{s,t}+\frac{m_0^2 G}{1+m_0^2 x G^2}
(W^-G_R(p))_{s,N_s}(G_R(p)W^+)_{N_s,t} \nn \\
M_+(p)_{s,t} &=& (W^- G_R(p)_{s,t}-\frac{m_0^2 x G}{1+m_0^2 x G^2}
(W^-G_R(p))_{s,N_s}G_R(p)_{N_s,t} \nn \\
M_-(p)_{s,t} &=& (W^+ G_L(p)_{s,t}-\frac{m_0^2 x G}{1+m_0^2 x G^2}
G_R(p)_{s,N_s}(G_R(p)W^+)_{N_s,t} \nn 
\end{eqnarray}
with $G \equiv G_R(p)_{N_s,N_s}$.
Similarly
\begin{eqnarray}
(D(p)^{-1}_{12})_{s,t} &=& 
\langle \psi(p)_s\psi(-p)_t\rangle \nn\\
&=&
\frac{m_0}{1+m_0^2 x G^2}
\Biggl[
x G_R(p)_{s,N_s}G_R(p)_{N_s,t} I_2 P_-
-i\ovl{p}_\mu\gamma_\mu  G_R(p)_{s,N_s}(G_R(p)W^+)_{N_s,t} I_2 P_+ \nn \\
& &+i\ovl{p}_\mu\gamma_\mu  (W^-G_R(p))_{s,N_s}G_R(p)_{N_s,t} I_2 P_-
+(W^-G_R(p))_{s,N_s}(G_R(p)W^+)_{N_s,t} I_2 P_+
\Biggr] 
\end{eqnarray}
and
\begin{eqnarray}
(D(p)^{-1}_{21})_{s,t} &=& 
\langle \bar\psi(p)_s\bar\psi(-p)_t\rangle \nn\\
&=&
\frac{m_0}{1+m_0^2 x G^2}
\Biggl[
x G_R(p)_{s,N_s}G_R(p)_{N_s,t} I_2 P_+
+i\ovl{p}_\mu\gamma_\mu^T 
G_R(p)_{s,N_s}(G_R(p)W^+)_{N_s,t} I_2 P_- \nn \\
& &-i\ovl{p}_\mu\gamma_\mu^T 
(W^-G_R(p))_{s,N_s}G_R(p)_{N_s,t} I_2 P_+
+(W^-G_R(p))_{s,N_s}(G_R(p)W^+)_{N_s,t} I_2 P_-
\Biggr] . 
\end{eqnarray}

\narrowtext
See ref.\cite{ANZ} for an application of this result.

\section*{Appendix C. Properties of Diagonalization Matrices}

In this appendix we derive several properties of diagonalization matrices,
$U$ and $V$, which are used for the renormalization of quarks fields.

Let us consider the tree level diagonalization of matrices
$ (W_0^\mp\cdot W_0^\pm)$.
To diagonalize  $ (W_0^\mp\cdot W_0^\pm)$, we have to solve the eigen-value 
problems $ (W_0^\mp\cdot W_0^\pm)_{s,t} \phi_\pm^i(t) = (M_0^2)_i 
\phi_\pm^i(s)$, then
$U_0$ and $V_0$ are given by normalized eigenvectors $\phi_\pm$:
$ (U_0)_{s,t} = \phi_+^s(t)$ and $(V_0)_{s,t}=\phi_-^s(t)$.
The two eigen-state equations lead to the same equation
\widetext
\begin{equation}
-w_0\left(\phi_\pm^i(s+1)+\phi_\pm^i(s-1)\right)+\left(1+w_0^2-(M_0^2)_i\right)
\phi_\pm^i(s) = 0
\end{equation}
\narrowtext
but with different boundary conditions:
\begin{equation}
-w_0 \phi_+^i (0) + \phi_+(1)=0, \qquad \phi_+^i(N_s+1) = 0
\end{equation}
or
\begin{equation}
-w_0 \phi_-^i (N_s+1) + \phi_-(N_s)=0, \qquad \phi_-^i(0) = 0 .
\end{equation}
Therefore, once $\phi_+^i(s)$ is known, the other is easily obtained
through $\phi_-^i(s) = \phi_+^i(N_s+1-s)$. Hereafter we consider
$\phi_+^i(s)$ only and drop the suffices $+$ and $i$.

There are two types of solutions to the eigen-state equation.
For $(M_0^2)_i \le (1-w_0)^2$ we have a damping solution $\phi(s) = 
A e^{-\alpha s}$ with $\cosh \alpha = \displaystyle
\frac{1+w_0^2-(M_0^2)_i}{2w_0}$. The first boundary 
condition leads to  $e^{-\alpha} = w_0$. This implies $(M_0^2)=0$, and 
therefore $\phi(s)$ is nothing but the zero mode solution
of the domain-wall QCD. For this solution $w_0$ should satisfy
$w_0^2\le 1$ ( $0\le M \le 2$).
The other boundary condition can be satisfied in the large $N_s$ limit.
The normalization constant becomes $A = (1-w_0^2)^{1/2}$.
Note that there are no other damping solutions which satisfy the first boundary
condition.

If the eigen-value is in the region $(1-w_0)^2 \le (M_0^2)_i
\le (1+w_0)^2$, we have an oscillating solution $\phi (s) =
A e^{i\alpha s}+ Be^{-i\alpha s} $ with $\cos \alpha = \displaystyle
\frac{1+w_0^2-(M_0^2)_i}{2w_0}$. The two boundary conditions imply
\begin{equation}
\left(
\begin{array}{cc}
e^{i\alpha}-w_0 & e^{-i\alpha}-w_0 \\
e^{i\alpha (N_s+1)} & e^{-i\alpha (N_s+1)} \\
\end{array}
\right)
\times
\left(
\begin{array}{c}
A \\
B \\
\end{array}
\right)  = 0 .
\end{equation}
The existence of the non-trivial solution requires
$ w_0 \sin\alpha (N_s+1) = \sin \alpha N_s $, which leads to
$\phi (s) = -A e^{i\alpha (N_s+1)} \sin \alpha (N_s+1-s)
\equiv A_0 \sin \alpha (N_s+1-s)$. Without loss of generality we
can take real $A_0$, and the normalization condition gives $A_0 =(2/N_s)^{1/2}
(1+O(1/N_s))$.
Setting $\alpha = a/N_s$ we reduce the equation for $\alpha$
to $w_0\sin a = \sin a$ in the large $N_s$ limit.
The solutions $a = \pi n$ with integer $n$ to this equation is
translated to $N_s -1$ independent solutions: $\alpha = \pi n /N_s$
with $n = 1,2,\cdots N_s-1$. (Note that $0\le \alpha_s\le \pi$
since $\sin \alpha_s > 0 $.) Therefore, all eigen-values and eigen-vectors
are now obtained, giving
\begin{equation}
\left[ U_0 \right]_{s,t} = \left\{
\begin{array}{cc}
(2/N_s)^{1/2}\sin \alpha_s (N_s+1-t) & s\not= N_s \\
(1-w_0^2)^{1/2} w_0^{(t-1)} & s = N_s \\
\end{array}
\right. ,
\end{equation}
and $ [V_0]_{s,t} = [U_0]_{s,N_s+1-t}$.

Next we prove some properties of $U_0$ and $V_0$.
It is noted that $U_0$ and $V_0$ can also diagonalize $W_0^\pm$ :
\begin{equation}
\left( V_0 W_0^+ U_0^\dagger\right)_{s,t}
=\delta_{s,t} f_s
\end{equation}
where 
\begin{equation}
f_s = \left\{
\begin{array}{cc}
w_0\cos\alpha_s(N_s+1) -\cos \alpha_s N_s & s\not= N_s \\
0 & s = N_s \\
\end{array}
\right. .
\end{equation}
Using the equation for $\alpha_s$ ($s\not= N_s$) we can show
\widetext
\begin{eqnarray}
f_s^2 &=& w_0^2+1 -2w_0 \left[\sin\alpha_s N_s\sin\alpha (N_s+1)+
\cos\alpha_s N_s \cos\alpha_s (N_s+1) \right] \nn \\
&=& w_0^2+1 -2w_0\cos\alpha_s = (M_0^2)_s .
\end{eqnarray}
This proves $f_s = (M_0)_s$ for all $s$.

It is also important to note that $U_0$ ($V_0$) diagonalizes
$I_{\rm log}^+$ ($I_{\rm log}^-$) terms, since
\begin{eqnarray}
\left( U_0 w_0^{s+t-2} U_0^\dagger\right)_{s,t}
&=& \delta_{s,t}\delta_{s,N_s}(1-w_0^2)\sum_{s,t}w_0^{2(s+t-2)} +O(1/N_s)
\nn \\
&=& \delta_{s,t}\delta_{s,N_s}(1-w_0^2)^{-1} + O(1/N_s) .
\end{eqnarray}

Now let us consider quantities including $g^2$ contributions.
We first show that $U$ and $V$ diagonalize $\ovl{W}^\pm$ at this order:
\begin{eqnarray}
V\cdot \ovl{W}^+\cdot U^\dagger &=&
(1+g^2 V_1)V_0(W_0^++g^2 W_1^+)U_0^\dagger (1+ g^2 U_1^\dagger) \nn \\
&=&V_0\cdot W_0^+ U_0^\dagger + g^2\left\{
V_1V_0W_0^+U_0^\dagger+V_0W_0U_0^\dagger U_1^\dagger+\wt{W_1}^+ 
\right\}
\label{eq:diag}
\end{eqnarray}
where the coefficient of the $g^2$ term is simplified to
\begin{eqnarray}
(V_1)_{s,t}(M_0)_t+(M_0)_s (U_1^\dagger)_{st} + (\wt{W_1^+})_{s,t}
&=&
(\wt{W_1^+})_{s,t} \left( 1+\frac{(M_0^2)_t-(M_0^2)_s}{(M_0^2)_s-(M_0^2)_t}
\right) = 0
\end{eqnarray}
\narrowtext
for $s\not= t$, and becomes
$ (\wt{W_1}^+)_{s,s}$ for $s=t$. Eq.(\ref{eq:diag}) then becomes
$$
= \left( M_0 + g^2\wt{W_1}^+ \right) \bf{1} .
$$
It is necessary for the stability of the zero mode to show that
$(\wt{W_1}^+)_{N_s,N_s} = 0$.
This can be proven as follows.
\begin{eqnarray}
(\wt{W_1}^+)_{N_s,N_s} &=& (1-w_0^2) \sum_{s,t} w_0^{N_s-s} (W_1^+)_{s,t}
w_0^{t-1}
\label{eq:zeroM}
\end{eqnarray}
where $W_1^+$ is composed of the sum $G_L$, $G_R$, $W_0^+G_L$ and
$W_0^-G_R$ but
$$
\sum_{s,t} w_0^{N_s-s} G(s,t) w_0^{t-1} = O(N_s w_0^{N_s}) \rightarrow 0
$$
for all $G=G_L$, $G_R$, $W_0^+G_L$ and $W_0^-G_R$ in the large $N_s$ limit. 
Therefore eq.(\ref{eq:zeroM}) vanishes.

For the wave-function renormalization factor we have to know
$$
U_{N_s,1} = V_{N_s,N_s} = (U_0)_{N_s,1} + g^2 \sum_{t\not= N_s}
(U_1)_{N_s,t}(U_0)_{t,1} .
$$
Fortunately, since $ (U_0)_{t,1} = (2/N_s)^{1/2} \sin\alpha_t (1-1) = 0$,
there is no order $g^2$ contribution and it becomes  
$U_{N_s,1}=(1-w_0^2)^{1/2}$.

Finally we would like to evaluate $ I_\pm^d$ from $I_{\rm finite}^\pm$.
If we define 
$$
\langle F(s,t) \rangle_U \equiv \sum_{s,t}(U_0)_{N_s,s} F(s,t)
(U_0)_{N_s,t}
$$
and
$$
\langle F(s,t) \rangle_V \equiv \sum_{s,t}(V_0)_{N_s,s} F(s,t)
(V_0)_{N_s,t} ,
$$
we can show
$$
\langle e^{-\alpha \vert s-t \vert} \rangle_{U,V}
=(1-w_0^2)
\frac{\sinh\alpha_0-\sinh\alpha}{2w_0\sinh\alpha_0 (\cosh\alpha_0-\cosh
\alpha )}
$$
with $e^{-\alpha_0} = w_0$,
$$
\langle e^{-\alpha (s+t-2)}\rangle_U = 
\langle e^{-\alpha (2N_s-s-t)}\rangle_V = (1-w_0^2)
\frac{e^{2\alpha}}{(e^\alpha-w_0)^2}
$$
and
$$
\langle e^{-\alpha (s+t-2)}\rangle_V = 
\langle e^{-\alpha (2N_s-s-t)}\rangle_U = 0 .
$$
Using these formula we obtain
\widetext
\begin{eqnarray}
\langle G_L(s,t)\rangle_U & = &
\frac{1-w_0^2}{2W\sinh\alpha}
\left[
\frac{\sinh\alpha_0 -\sinh\alpha}{2w_0\sinh\alpha_0 (\cosh\alpha_0
-\cosh\alpha)}
-\frac{e^\alpha-W}{e^{-\alpha}-W}\frac{1}{(e^\alpha-w_0)^2}
\right] \nn \\
& = & \langle G_R(s,t)\rangle_V \equiv \wt{G_L} \\
\langle G_R(s,t)\rangle_U & = &
\frac{1-w_0^2}{2W\sinh\alpha}
\left[
\frac{\sinh\alpha_0 -\sinh\alpha}{2w_0\sinh\alpha_0 (\cosh\alpha_0
-\cosh\alpha)}
-\frac{1}{(e^\alpha-w_0)^2}
\right] \nn \\
& = & \langle G_L(s,t)\rangle_V \equiv\wt{G_R}
\end{eqnarray}
\narrowtext
and
$$
 \langle W_0^+ G_L(s,t) \rangle_{U,V}
= \langle W_0^- G_R(s,t) \rangle_{U,V}
= (w_0-W)\wt{G_R} .
$$
The explicit expression for $I_{\rm finite}^\pm$ is reduced to
the final result in terms of $\wt{G}_{L/R}$: 
$I_+^d = I_-^d \equiv I^d$ where
\widetext
\begin{eqnarray}
I^d & = & C_2 \int\frac{d^4l}{(2\pi)^4}\left\{
\frac{1}{8\hat l^2} \sum_\mu \left[  \sin^2l_\mu ( \wt{G_R}+\wt{G_L})
+ 2 \cos l_\mu (w_0-W(l))\wt{G_R} \right] \right. \nn \\
&+& \sum_\mu \frac{\sin^2 l_\mu}{2(\hat l^2)^2} 
\left[(w_0-W(l))\wt{G_R}+(\sum_\nu \cos^2 l_\nu/2 - 2\cos^2l_\mu/2 )
\wt{G_L}+\sum_\nu (\sin^2l_\nu/2) \wt{G_R} \right] \nn \\
&-&\left. \frac{1}{(l^2)^2}\theta(\pi^2-l^2)\right\} .
\end{eqnarray}
\narrowtext

\newcommand{\J}[4]{{\it #1} {\bf #2} (19#3) #4}
\newcommand{\MPL}{Mod.~Phys.~Lett.}
\newcommand{\IJMP}{Int.~J.~Mod.~Phys.}
\newcommand{\NP}{Nucl.~Phys.}
\newcommand{\PL}{Phys.~Lett.}
\newcommand{\PR}{Phys.~Rev.}
\newcommand{\PRL}{Phys.~Rev.~Lett.}
\newcommand{\AP}{Ann.~Phys.}
\newcommand{\CMP}{Commun.~Math.~Phys.}
\newcommand{\PTP}{Prog. Theor. Phys.}
\newcommand{\Suppl}{Prog. Theor. Phys. Suppl.}

\newpage

\newpage

\begin{table}
\caption{Value of $I^d$ vs $\wt{M}$, together with $Z_1$ and
$(Z_1)_{\rm non-tad}/Z_1$}
\label{tab:wave}
\begin{center}
\begin{tabular}{llll}
$\wt{M}$ & $I^d$ & $Z_1 $ & $(Z_1)_{\rm non-tad}/Z_1$ \\
\hline
0.05 & -0.01945(5) & -0.09897 &  0.041 \\
0.10 & -0.01871(5) & -0.09822 &  0.049 \\
0.15 & -0.01804(5) & -0.09756 &  0.056 \\
0.20 & -0.01744(5) & -0.09696 &  0.063 \\
0.25 & -0.01688(5) & -0.09640 &  0.069 \\
0.30 & -0.01636(5) & -0.09589 &  0.075 \\
0.35 & -0.01588(5) & -0.09541 &  0.080 \\
0.40 & -0.01544(5) & -0.09496 &  0.085 \\
0.45 & -0.01502(5) & -0.09454 &  0.090 \\
0.50 & -0.01463(5) & -0.09415 &  0.095 \\
0.55 & -0.01426(5) & -0.09378 &  0.099 \\
0.60 & -0.01392(5) & -0.09345 &  0.103 \\
0.65 & -0.01361(5) & -0.09313 &  0.107 \\
0.70 & -0.01332(5) & -0.09284 &  0.110 \\
0.75 & -0.01305(5) & -0.09257 &  0.113 \\
0.80 & -0.01281(5) & -0.09233 &  0.116 \\
0.85 & -0.01259(5) & -0.09211 &  0.119 \\
0.90 & -0.01239(5) & -0.09191 &  0.121 \\
0.95 & -0.01222(5) & -0.09174 &  0.124 \\
\end{tabular}
\end{center}
\end{table}

\end{document}